\documentclass{epl}
\title{ Competition Between Exchange and Anisotropy in a \\
Pyrochlore Ferromagnet}
\author{ J. D. M. Champion \inst{1,2,3} \and S. T. Bramwell \inst{1} \and
P. C. W. Holdsworth \inst{2} \and M. J. Harris \inst{3} }
\institute{
  \inst{1} University College London, Department of Chemistry, 20 Gordon
Street, London WC1H0AJ, U.K.\\
  \inst{2} Laboratoire de Physique, Ecole Normale Sup\'erieure, 46 All\'ee
d'Italie, F69634 Lyon cedex 07, France.\\
\inst{3} ISIS Facility, Rutherford Appleton Laboratory, Chilton, Didcot,
Oxon. OX11 0QX, U.K.
}
\pacs{75.10.Hk}{Classical spin models}
\pacs{75.40.Mg}{Numerical simulation studies}
\pacs{75.30.Kz}{Magnetic phase boundaries}

\begin{document}

\maketitle

\begin{abstract}
The Ising-like spin ice model, with a macroscopically degenerate ground
state, has been shown to be approximated by several real materials. Here we
investigate a model related to spin ice, in which the Ising spins are
replaced by classical Heisenberg spins. These populate a cubic pyrochlore
lattice and are coupled to nearest neighbours by a ferromagnetic exchange
term $J$ and to the local $\langle 1,1,1 \rangle$ axes by a single-ion
anisotropy term $D$. The near neighbour spin ice model corresponds to the
case $D/J \rightarrow \infty$. For finite $D/J$ we find that the
macroscopic degeneracy of spin ice is broken and the ground state is
magnetically ordered into a four-sublattice structure. The transition to
this state is first-order for $D/J > 5$ and second-order for $D/J < 5$ with
the two regions separated by a tricritical point. We investigate the
magnetic phase diagram with an applied field along $[1,0,0]$ and show that
it can be considered analogous to that of a ferroelectric.
\end{abstract}

\section{Background}
Geometrical frustration represents a recipe by which condensed matter can
be disordered even in the absence of substitutional disorder \cite{Ziman}.
The canonical example is the proton disorder in ice, which was famously
shown by Pauling to be the origin of the experimentally observed  ground
state entropy \cite{Pauling,GS}. Anderson later illustrated a direct mapping
of Pauling's model onto the Ising  antiferromagnet on the pyrochlore
lattice \cite{Anderson}. However, Anderson's antiferromagnet does not
appear to occur in nature and the mapping was for many years something of
an academic curiosity, albeit one that inspired interest in frustrated
magnetism \cite{Greedan,Moessner,Shender}. Recently, however, this has
changed with the presentation of an alternative mapping named ``spin ice'',
motivated by
experiments on
the cubic pyrochlore Ho$_2$Ti$_2$O$_7$ \cite{PRL1}. This material is
closely approximated by a
pyrochlore {\it ferromagnet} with $\langle 1,1,1 \rangle$ Ising-like spins,
which maps onto the ice model \cite{PRL1} and hence onto Anderson's
antiferromagnet \cite{JPCM,Roderich}. The behaviour of Ho$_2$Ti$_2$O$_7$
\cite{PRL1}
and the related compound Dy$_2$Ti$_2$O$_7$ \cite{PRL2,Ramirez} is described
to a good first approximation by
the spin ice model and has recently been shown to be reproduced to an
excellent degree by a related
``dipolar spin ice model'' including long range dipole
interactions\cite{Byron1,Byron2,SQ}.
These materials are part of a series of rare earth titanates that are in
general of interest for their frustrated magnetic behaviour
\cite{JMMM,titanate}.

\begin{figure}
\onefigure[scale=0.3]{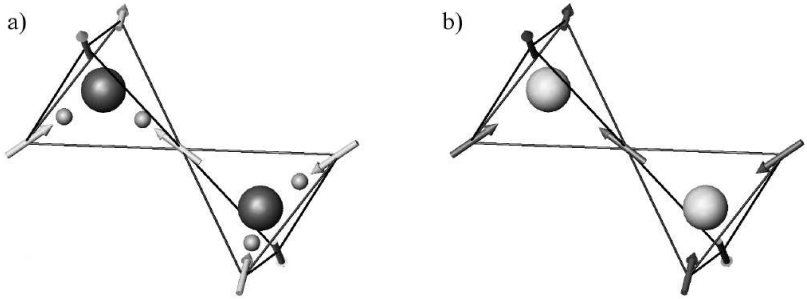}
\caption{The spin ice mapping \cite{PRL1,JPCM}. a) Coordination of H$_2$O
molecules in ice represented by proton displacement vectors. b) Spins in
the spin ice model.
}
\label{f1}
\end{figure}

The mapping  may be illustrated as follows \cite{PRL1} (see Fig. \ref{f1}).
In Ho$_2$Ti$_2$O$_7$ and Dy$_2$Ti$_2$O$_7$ the magnetic rare earth
ions occupy a cubic pyrochlore lattice, an array of corner-linked
tetrahedra.  This lattice is the medial lattice (i.e. the lattice formed by
the mid-points of the bonds) of
the diamond-type oxide sublattice in cubic ice \cite{Anderson}.
The proton disorder of ice can be described by
displacement vectors that populate the vertices of this lattice. The
``ice rules'', the condition that
there are two protons near to, and two further away from each oxide,
corresponds to the condition
that two vectors point into and two point out of each tetrahedron. In
Ho$_2$Ti$_2$O$_7$ and Dy$_2$Ti$_2$O$_7$ the rare earth ground state is an
effective S $= 1/2$ doublet with local $\langle 1, 1, 1 \rangle$
quantization axis and so can be
represented by a classical Ising-like vector analogous
to a proton displacement vector.  Net ferromagnetic coupling between near
neighbour spins completes the analogy
with the ice model, corresponding to the frustrated proton-proton repulsion
imposed by the ice structure.
The spin ice model describes qualitatively the results of many experiments
on these materials,
including the absence of magnetic order down to the lowest temperatures
\cite{PRL1, JMMM, Ramirez}.

\section{The Model}

The macroscopic degeneracy of the spin ice model makes it very susceptible
to perturbations. That is, one
can expect correction terms to the idealized Hamiltonian to dramatically
change the low temperature
properties by lifting the degeneracy, moving the system into a long range
ordered state. This is the
case for dipolar interactions, as long as thermodynamic equilibrium is
maintained.
In this paper we pursue the effects of perturbation by studying a
continuous spin ice model, with
classical Heisenberg spins replacing the discrete Ising-like degrees of
freedom.
Our model is defined by the spin Hamiltonian
\begin{equation} \label{Ham}
H = - J \sum_{<i,j>} \vec S_i.\vec S_j -D \sum_i (\vec S_i.\vec d_i)^2
- \sum_i \vec H.\vec S_i,
\end{equation}

\noindent
where the $\vec S_i$ are classical vectors of unit length.
The cubic crystal fields of magnitude $D$ are along
the four $\langle 1,1,1 \rangle$
directions for the four spins of the primitive unit cell.
The extra refinement of continuous, rather than Ising spins
would be expected to capture the effect of quantum fluctuations, which must
be present
in some extent in all real materials. One would also expect it to be an
accurate
starting Hamiltonian for any prospective ice or spin ice like materials
with an exchange constant
$J$ in excess of the $O(1^o)$ $K$ temperature range of Ho$_2$Ti$_2$O$_7$
and Dy$_2$Ti$_2$O$_7$,
where dipolar interactions are important. The
Hamiltonian (\ref{Ham}) describes the transition from spin ice ($D/J
\rightarrow \infty$)
to ordinary ferromagnetic ($D/J \rightarrow 0$) behaviour.  We have
investigated it by Monte Carlo simulation using the standard Metropolis
spin-flipping algorithm for systems of size $N = 432 - 11634$ spins. All
simulation lengths were 100,000 Monte Carlo steps per spin with 30-40,000
steps used for initial equilibration.

\section{Spontaneous Magnetization}
\begin{figure}
\onefigure[scale=0.4]{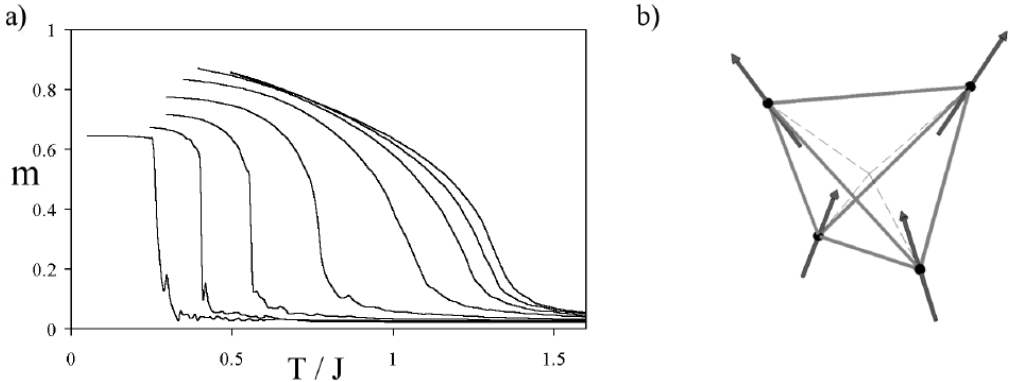}
\caption{a) Monte Carlo simulation of the magnetization versus temperature
for Hamiltonian \ref{Ham}. Thermal hysteresis was observed for large $D/J$;
the results shown are cooling curves at (from left to right) $D/J = 22.91,
15.49, 10.47, 7.08, 4.79, 3.24, 2.18, 1.0$. b) Spin structure of the $q =
0$ basis for $D/J = 7$, illustrating the canting of the spins towards
$[0,0,1]$.} direction.
\label{f2}
\end{figure}

Our first result is illustrated in Fig. \ref{f2}, where we plot Monte Carlo
simulation results of the scalar magnetization per
spin $m = \frac{1}{N} \left\langle \sqrt{\sum_i (\vec{S_i})^2} \right\rangle$  as a
function
of temperature $T/J$ for $0 <  D/J < 25$. For all the finite $D/J$
investigated there is a transition to an ordered
state with $q=[0,0,0]$ propagation vector.
It is clear
that the transition only disappears in the spin ice limit, $D/J \rightarrow
\infty$ and therefore that the infinite ground state degeneracy is lifted,
for all finite $D/J$,
in favour of a long range ordered $q=0$ state.
The transition is marked by a maximum in specific heat and
susceptibility. It is second-order for $D = 0$
(the simple ferromagnetic case) but clearly first-order for large $D/J$,
suggesting the presence of a tricritical point \cite{tri1,tri2} at
intermediate $D/J$. To
locate the tricritical point $D_{tc}/J$ we have performed a finite size
scaling analysis of the maximum in the susceptibility $\chi_{max}$.
 $\chi_{max}$ should scale
as $L^{\gamma/\nu}$ in the second-order regime \cite{FSS}, where $\gamma$
and $\nu$ are the critical
exponents describing the divergence of the susceptibility and correlation
length.
Assuming the Ising universality class, one has $\gamma/\nu = 1.969 \approx
2$ \cite{Binney}.
Such $L^2$ scaling was observed for
$D/J < 5$ only, and a careful analysis suggests that the tricritical point
occurs at $D_{tc}/J = 5.0 \pm 0.5$.

To further understand the results of Fig. \ref{f2}, consider the nature of the
ordered state: a $q=0$ state has a magnetic unit cell consisting of a single
tetrahedron.
For the Ising system ($D/J \rightarrow \infty$)
the spins are directed along the local $\langle 1,1,1 \rangle$ axes, 
as shown in Fig. \ref{f1}b. We call this the ``$\langle 1,1,1
\rangle$'' state which has energy per spin $E_{\langle 1,1,1 \rangle}
= -J/3 -D$ and overall
magnetization $m = 1/\sqrt{3}$ along $\vec z  = [0,0,1]$ . The state is
frustrated by the crystal field with, for each tetrahedron, four bonds
making
a contribution of $-J/3$ and two making a contribution of $J/3$.
Introduction of continuous spins allows the
system to regain
frustrated bond energy at the expense of field energy by relaxing away from
the body centered cubic directions; see Fig. \ref{f2}b.
In disordered spin ice ground states this relaxation is itself frustrated
and it is
easy to see that the maximum energy
gain will occur for relaxation away from the ordered $q=0$ state.
For finite $D/J$ we therefore expect degeneracy to be lifted in favour of
this state and to have
a canting of the  spins towards the $[0,0,1]$. In
the limit $D/J \rightarrow 0$ one should obtain a
perfectly aligned ferromagnetic (FM) state, which we call the ``$[0,0,1]$
state''. For finite $D$ it has
energy $E_{[0,0,1]} = -3J - D/3$.

We have estimated the ground state spin orientation for large and small
crystal field
by making the reasonable ansatz that the relaxation
away from, or towards the $\langle 1,1,1 \rangle$ is homogeneous for the
four spins of
the unit cell. Comparing with numerical data this seems {\it a posteriori}
to be correct.
For small $\alpha = D/J$ a small relaxation away from $[0,0,1]$ and towards
$\langle 1,1,1 \rangle$ gives for the  magnetization $m$ and the energy $E$
per spin
\begin{equation}\label{res-1}
 E =J \left( -3 -{\alpha\over{3}} - {\alpha^2\over{18}}\right), \;\;
 m = \sqrt{\left(1-{\alpha^2\over{72}}\right)}.
\end{equation}
This result illustrates the fact that the energy and magnetization
change very slowly from the ferromagnetic value for $D$ increasing from zero
to $D \sim O(J)$. In the opposite limit of strong crystal field, we define
$\gamma = J/D$ and allow a
small relaxation away from the $\langle 1,1,1 \rangle$
state towards the $[0,0,1]$ state. In this case we find for small $\gamma$
\begin{equation} \label{res-2}
E = D\left(-1 - {\gamma\over{3}} - {32\gamma^2\over{9}}\right), \;\;
m = {1\over{\sqrt{3}}} + {8\gamma\over{3\sqrt{3}}},
\end{equation}
which again suggests a slow change in magnetization with $D/J$.

\begin{figure}
\onefigure[scale=0.3]{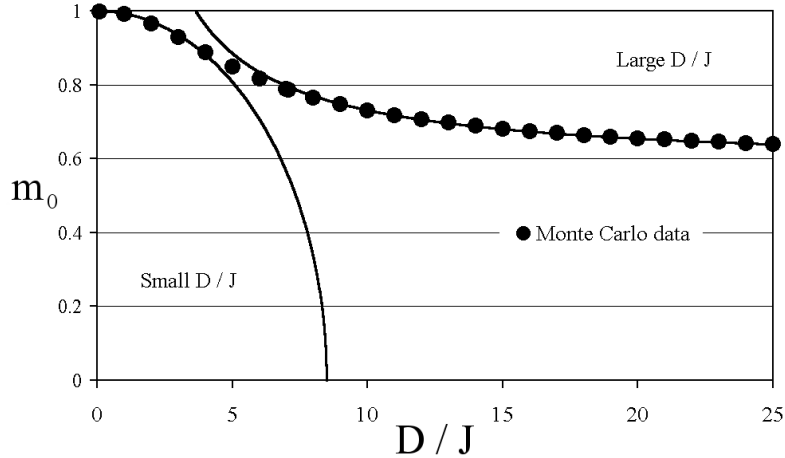}
\caption{Comparison of the numerical zero temperature magnetization
$m_0$, estimated by a zero-temperature Monte Carlo method,
with that estimated by the analytical expressions (\ref{res-1}) (left hand
line) and (\ref{res-2}) (right hand line). Both asymptotic expressions are
seen to break down near the tricritical point $D_{tc}/J = 5.0 \pm 0.5$.}
\label{f3}
\end{figure}

To test these expressions, the magnetization was computed by performing
zero temperature Monte Carlo simulations.
In Fig. \ref{f3} we plot these estimates versus the anisotropy $D/J$ and
compare  with the small anisotropy result (\ref{res-1})
and the large anisotropy result (\ref{res-2}).  The agreement is seen to be
excellent in both cases, with each asymptotic equation breaking down near the
tricritical point at $D_{tc}/J = 5$. It is therefore tempting to associate the
change in order of the transition with a non-linear cross over from a regime
that is nearly ferromagnetic, to one that is nearly $q = 0$ spin ice.

\section{Magnetic Phase Diagram}
Further insight into the nature of this cross over is gained by a
consideration of the magnetic phase diagram.
We consider the behaviour of the continuous spin ice model in an external
magnetic field applied along the
$[1,0,0]$ direction which sustains the symmetry of the ordered state for
all finite $D/J$.  In ref.(~\cite{PRL2})
we showed that the magnetic phase
diagram of the near neighbour spin ice model with the field along $[1,0,0]$
has two lines
of first-order phase transitions that separate phases of different
magnetization. Both lines terminate in critical points, and we have argued
that
the phase diagram can be considered analagous to that of a liquid-gas system,
where a line of first-order transitions separates phases of different density.
For the continuous spin ice model we have mapped out the equivalent phase
diagram by Monte Carlo simulation. In each case the phase boundary was
estimated
from the maximum in the magnetic susceptibility. In principle, now we know
the true
ground state, we could measure the true order parameter and its
fluctuations,  by calculating
the projection onto the calculated state.

\begin{figure}
\onefigure[scale=0.5]{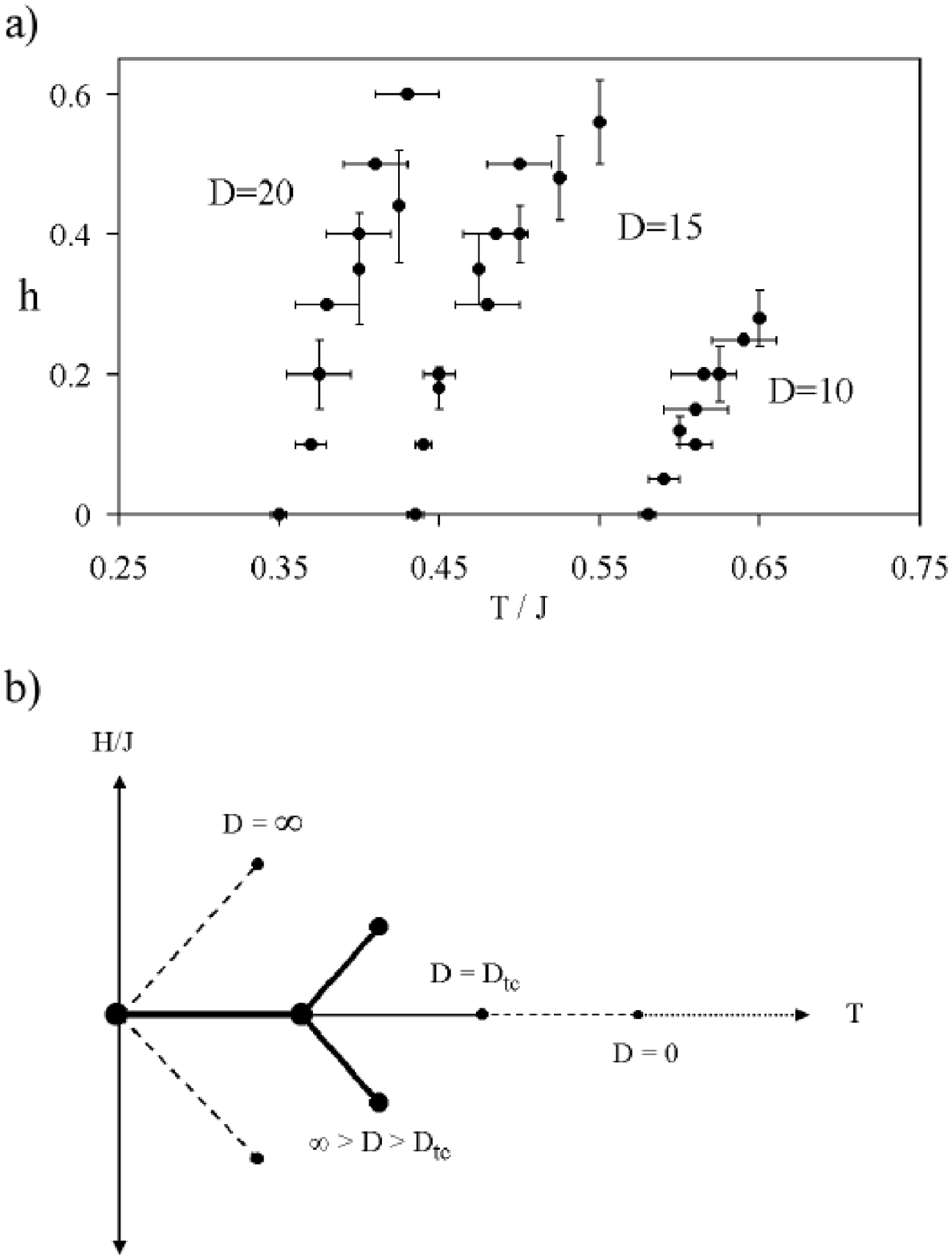}
\caption{Magnetic phase diagram for the continuous spin ice model define by
eqn. (\ref{Ham}). (a) Phase diagram determined by Monte Carlo simulation on
a system of $N = 2000$ spins. Points on the coexistence lines were
determined from the maximum in the susceptibility in fixed temperature
(vertical error bars) or fixed field (horizontal error bars) Monte Carlo
runs. For ease of comparison with experiment and ref.(\cite{PRL2}), the
field $h$ is defined as $h = H \times (\mu_{\rm B}/k_{\rm B} = H/0.06717$,
with $\mu_{\rm B}$ and $k_{\rm B}$ the Bohr magneton and Boltzmann's
constant respectively, and $H$ defined in eqn.\ref{Ham}.  (b) Schematic
phase diagram showing the case of $D/J = 0$ (short dashed line), $D/J =
D_{tc}/J$ (thin line), $D/J > D_{tc}/J $ (thick line) and $D/J \rightarrow
\infty$ (long dashed line)}
\label{f4}
\end{figure}

The results are shown in Fig. \ref{f4}a, and represented schematically in
Fig. \ref{f4}b. It is
seen that the two first-order lines
of the spin ice phase diagram coalesce below the ordering temperature,
which is therefore seen to be a triple point (see Fig. \ref{f4}b).
The triple point temperature increases with
decreasing $D/J$ and eventually the two ``wings'' disappear and give way to
a line
of first-order transitions along the zero field axis, typical of a
ferromagnet.
We would anticipate that the two wings disappear precisely at the
tricritical point determined above, $D_{tc}/J \approx 5$. This kind of
phase diagram is exhibited by idealised ferroelectrics and also some real
ones such as BaTiO$_3$ \cite{ferroelectrics}. It can be rationalised by
Landau theory \cite{ferroelectrics,Aharony}, where the free energy is
written
$G = G_0 + \frac{c_2}{2} m^2 + \frac{c_4}{4} m^4 + \frac{c_6}{6} m^6 -
NmH$. Here, $m$ and $H$  are the
order parameter and conjugate applied field and the $c_n$'s are constants
related to the n-th order susceptibilities. If $c_4 > 0$ then the
zero-field transition is second-order and if $c_4 < 0$ it is first-order,
with the characteristic ``winged'' phase diagram. The slopes of the wings
are approximately $\sim \sqrt{\frac{c_4}{c_2}}$. In general, termination
points of first-order lines need not be associated with critical
fluctuations, in which case they should be termed ``first-order'' critical
points. One such point is known to occur on the zero-field axis of the
phase diagram of the one dimensional Ising ferromagnet with additional
inverse square interaction \cite{Thouless}. In fact it has been suggested
\cite{Aharony} that the end-point of a line of symmetry sustaining
transitions is usually first-order, the  liquid gas transition being a
special case. In general, at a first-order critical point, the
susceptibility will diverge, but there may or may not be critical
fluctuations \cite{Aharony,FB}). A detailed analysis of this question with
regard to the current system would be an interesting topic of future study.

\section{Conclusions}
In conclusion, it is noteworthy that the continuous spin ice model displays
behaviour characteristic of ferroelectrics. It is plausible that ice itself
would exhibit such a phase diagram if the dynamics did not become
immeasurably slow at temperatures well above the triple point \cite{GS}.
However ice is not proton ordered in the absence of
defects or stabilizing surfaces and there is much debate concerning the
true nature of its ground state \cite{ice0,ice1,ice2,ice3,ice4}.
Very similar remarks apply to Ho$_2$Ti$_2$O$_7$, where the dynamics also
become immeasurably slow below $0.7$ K. In view of our results, it might be
that the ordered state favoured by the long range part of the dipolar
coupling competes with that favoured by continuous spins. Through the
application of a magnetic field in the
$[1,1,0]$ direction \cite{PRL1}, it is possible to put Ho$_2$Ti$_2$O$_7$
into an ordered state of the same symmetry as the ground state of the
dipolar spin ice model \cite{Byron2}. It would be interesting to prepare a
sample in this way at a temperature below the expected ordering transition
at $0.18$ K, and to examine the stability of the ordered state as the field
is removed. At a theoretical level, it would be extremely interesting to
consider the effect of continuous spins on the dipolar spin ice model of
den Hertog and Gingras \cite{Byron1}, that is an accurate description of
Ho$_2$Ti$_2$O$_7$ and Dy$_2$Ti$_2$O$_7$ \cite{Byron1,SQ}.



\acknowledgments
It is a pleasure to thank J.T. Chalker and M.J.P. Gingras for stimulating
discussions. JDMC thanks the ENS, the EPSRC and ISIS for financial support.
This work was supported by the P\^ole Scientifique de Mod\'elisation
Num\'erique at the \'Ecole Normale Sup\'erieure de Lyon.

\end{document}